\documentclass{article} 
\usepackage{iclr2025_conference,times}
\iclrfinalcopy 

\usepackage{amsmath,amsfonts,bm}

\def\eqref#1{equation~\ref{#1}}

\def\1{\bm{1}}

\DeclareMathAlphabet{\mathsfit}{\encodingdefault}{\sfdefault}{m}{sl}
\SetMathAlphabet{\mathsfit}{bold}{\encodingdefault}{\sfdefault}{bx}{n}

\usepackage{hyperref}
\usepackage{url}
\usepackage{enumitem}
\usepackage{xspace}
\usepackage{graphicx}
\usepackage{cleveref}
\usepackage{wrapfig}
\usepackage{booktabs}
\usepackage{wrapfig}
\usepackage{booktabs}
\usepackage{subcaption} 

\usepackage{listings}
\usepackage{widetable}
\usepackage{upquote}
\usepackage[ruled,vlined]{algorithm2e}
\usepackage{multirow}

\newcommand{\WholeTitle}{WorkflowLLM\xspace}
\newcommand{\BenchTitle}{WorkflowBench\xspace}
\newcommand{\ModelName}{WorkflowLlama\xspace}

\newcommand{\SeedNum}{$14,771$\xspace}
\newcommand{\APINum}{$1,503$\xspace}
\newcommand{\APPNum}{$83$\xspace}
\newcommand{\CategoryNum}{$28$\xspace}
\newcommand{\TotalNum}{$106,763$\xspace}

\crefname{section}{§}{§§}
\Crefname{section}{§}{§§}

\title{WorkflowLLM: Enhancing Workflow Orchestration Capability of Large Language Models}

\author{Shengda Fan$^{1}\thanks{\ \ Indicates equal contribution.}$\hspace{0.5em}, Xin Cong$^{2*}$\thanks{\ \  Corresponding author.}\ \ , Yuepeng Fu$^2$, Zhong Zhang$^{2}$, Shuyan Zhang$^{3}$,  Yuanwei Liu$^4$,\\\textbf{ Yesai Wu$^2$, Yankai Lin$^{1\dag}$, Zhiyuan Liu$^{2}$, Maosong Sun$^2$}   \\
$^1$Renmin University of China
$^2$Tsinghua University 
$^3$The University of Manchester \\
$^4$Wuhan University
\\
\texttt{fanshengda@ruc.edu.cn, xin.cong@outlook.com}\\
}

\begin{document}

\maketitle

\begin{abstract}

Recent advancements in large language models~(LLMs) have driven a revolutionary paradigm shift in process automation from Robotic Process Automation to Agentic Process Automation by automating the workflow orchestration procedure based on LLMs. 
However, existing LLMs (even the advanced OpenAI GPT-4o) are confined to achieving satisfactory capability in workflow orchestration. 
To address this limitation, we present \WholeTitle, a data-centric framework elaborately designed to enhance the capability of LLMs in workflow orchestration.
It first constructs a large-scale fine-tuning dataset \BenchTitle with \TotalNum samples, covering \APINum APIs from \APPNum applications across \CategoryNum categories.
Specifically, the construction process can be divided into three phases:
(1) Data Collection: we collect real-world workflow data from Apple Shortcuts and RoutineHub, transcribing them into Python-style code.
We further equip them with generated hierarchical thought via ChatGPT.
(2) Query Expansion: we prompt ChatGPT to generate more task queries to enrich the diversity and complexity of workflows.
(3) Workflow Generation: we leverage an annotator model trained on collected data to generate workflows for synthesized queries. Finally, we merge the synthetic samples that pass quality confirmation with the collected samples to obtain the \BenchTitle.
Based on \BenchTitle, we fine-tune Llama-3.1-8B to obtain \ModelName.
Our experiments show that \ModelName demonstrates a strong capacity to orchestrate complex workflows, while also achieving notable generalization performance on previously unseen APIs. Additionally, \BenchTitle exhibits robust zero-shot generalization capabilities on an out-of-distribution task planning dataset, T-Eval. 
Our data and code are available at \url{https://github.com/OpenBMB/WorkflowLLM}.

\end{abstract}

\section{Introduction}

%

Process Automation~(PA)~\citep{cichocki1997workflow}, as a long-standing pursuit of the human race, aims to automate repetitive tasks to minimize human labor and improve efficiency. 
Tracing back to the agricultural era, humanity has employed waterwheels and oxen to automate farming practices.
Robotic Process Automation~(RPA), the current predominant PA technique, abstracts the repetitive task into a workflow (i.e., a program that can execute automatically) by orchestrating various actions (e.g., functions or APIs)~\citep{ivanvcic2019robotic,hofmann2020robotic,wewerka2020robotic,agostinelli2020towards,ferreira2020evaluation}.
While RPA successfully reduces the human labor via automated workflow execution, the process of orchestrating workflows still requires substantial manual effort. 
Recently, large language models~(LLMs)~\citep{openaichatgptblog,openai2023gpt4,touvron2023llama,touvron2023llama2,dubey2024llama} have achieved remarkable performance beyond natural language processing~\citep{ahn2022can,cheng2023gpt,DBLP:conf/acl/QianLLCDL0CSCXL24}.
The emergence of LLMs has unveiled a paradigm shift trend, moving from Robotic Process Automation to Agentic Process Automation~(APA)~\citep{ye2023proagent,zeng2023flowmind,huang2024promptrpa,wornow2024automating,li2024autoflow} which automates the workflow orchestration process by utilizing LLMs to build the workflow.

However, such a paradigm shift trend is constrained by \textbf{the limited ability of LLMs to orchestrate complex workflows}, which in turn leads to two crucial limitations in current APA methods:
(1) \textbf{Constrained Action Scale}: Current LLMs can only orchestrate small-scale workflows with a limited number of actions. The most advanced OpenAI GPT-4 is capable of managing workflows with an average of only 6.1 actions, even when equipped with advanced decision-making mechanisms \citep{ye2023proagent}. This falls short of the complexity required to meet real-world demands.
For instance, as a widely-used representative, Apple Shortcuts~\citep{apple2024shortcuts} involves an average of $70.4$ actions. 
(2) \textbf{Simple Logical Structure}: Currently, most existing work mainly focuses on generating sequential actions~\citep{yao2022react,DBLP:conf/iclr/QinLYZYLLCTQZHT24,chen2024t} while workflows of the real-world applications usually involve intricate logical structures such as branches and loops. 
For example, Apple Shortcuts averages $2.6$ nested branch/loop logical structures.
As a result, \textbf{there is an urgent need to unlock the workflow orchestration capability of LLMs to expedite the paradigm shift in process automation.}

\begin{figure}[!t]
    \centering
    \includegraphics[width=0.9\linewidth]{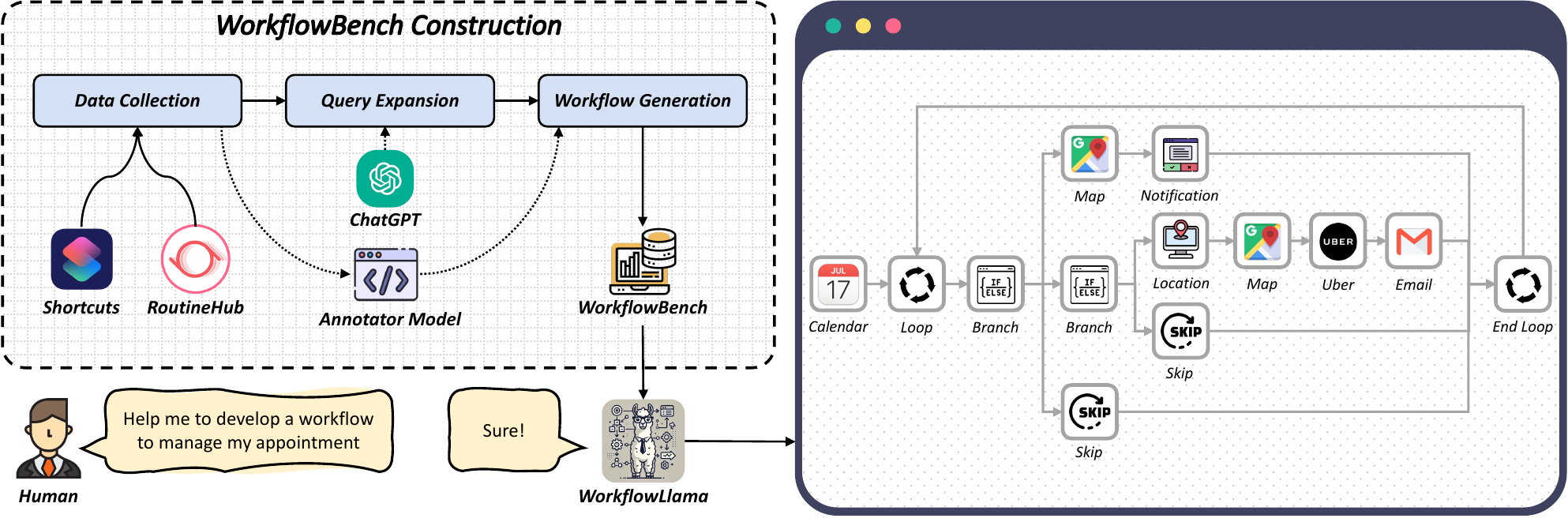}
    \caption{Overview of \WholeTitle. It first constructs \BenchTitle through a three-phase pipeline and fine-tunes \ModelName, which can generate workflows based on the user's query (appointment management in this case).}
    \label{fig:intro}
\end{figure}

To address these challenges, we propose \textbf{\WholeTitle}, a data-centric framework including dataset construction, model training, and evaluation to enhance LLMs' workflow orchestration capabilities (shown in \Cref{fig:intro}).
Specifically, we first construct \BenchTitle, which consists of \TotalNum supervised fine-tuning instances, encompassing \APINum APIs across \APPNum applications, structured through three primary phases:%
\begin{itemize}[topsep=-1pt, partopsep=1pt, leftmargin=12pt, itemsep=-1pt]
    \item \textbf{Data Collection}: 
    We select shortcuts from RoutineHub as high-quality data sources because they represent a robust RPA application with numerous expert-developed workflows available.
    We curate \SeedNum human-annotated, high-quality shortcuts spanning \CategoryNum diverse categories (e.g., Business, Health \& Fitness, Productivity), alongside associated metadata including titles, functionality descriptions, and API documentations.
    As the raw workflow data is not directly suitable for LLMs to process, and considering that Python allows more convenient parameter passing and control logic~\citep{ye2023proagent, wangexecutable}, we transcribe the shortcut source code into Python-like code.
    Subsequently, we prompt ChatGPT to generate comments, task plans, and task queries at varying levels of granularity—from fine-grained to coarse-grained—to enrich the data with detailed thought processes and enhance the learning efficacy of LLMs~\citep{wei2023chainofthought}.

    \item \textbf{Query Expansion}:
    To enrich the diversity and complexity of workflows, we utilize ChatGPT to generate additional task queries.
    Specifically, we first sample applications with diverse functionalities and select their APIs, along with built-in APIs, to prompt ChatGPT to generate task queries that leverage these sampled APIs to accomplish specific tasks.
    To further ensure workflow complexity, we also sample real-world workflow examples as demonstrations to guide ChatGPT in generating similar workflows.

    \item \textbf{Workflow Generation}:
    As existing LLMs even GPT-4o still struggle in workflow generation, we first train a workflow annotator model based on the collected real-world shortcuts.
    Then we utilize the trained annotator to generate workflows for the expanded task queries.
    To prevent low-quality workflows generated by the annotator model from affecting subsequent training, we perform quality confirmation to ensure dataset integrity.
    We first utilize ChatGPT to refine the generated workflows to fix existing minor bugs in them and then use rule-based filtering to remove workflows with logical errors.
\end{itemize}

To evaluate the capability of LLMs in workflow orchestration, we employ two metrics: the reference-code-based metric \textbf{CodeBLEU} and the model-based metric \textbf{Pass Rate}. 
Experimental results demonstrate that \ModelName consistently and significantly outperforms all baselines, including GPT-4o even with the in-context learning technique, across both metrics under unseen instructions and unseen APIs settings.
Furthermore, \BenchTitle demonstrates strong generalization capabilities in out-of-distribution (OOD) scenarios, particularly on the T-Eval benchmark~\citep{chen2024t}, where it achieves an F1 plan score of \textbf{77.5}\%.

\section{Related Work}

\paragraph{Process Automation}
RPA has gained considerable attention for automating repetitive tasks in various productivity scenarios~\citep{ivanvcic2019robotic,hofmann2020robotic,wewerka2020robotic,agostinelli2020towards,ferreira2020evaluation}.
%
RPA predominantly relies on handcrafted workflows (e.g., programming, recording human behavior), making them highly suitable for automating well-structured, routine processes~\citep{herm2020consolidated}. 
However, such approaches require substantial efforts and in-depth domain expertise, resulting in high setup costs and limited adaptability.
Recent advancements in LLMs have spurred interest in integrating these models into RPA to enhance flexibility and reduce dependency on manual workflow creation.
\citet{ye2023proagent} introduced the concept of APA, which utilizes LLMs to autonomously orchestrate workflows based on human instructions. 
Subsequently, several studies have sought to apply APA in various domains, including travel planning~\citep{xie2024travelplanner}, smartphone applications~\citep{huang2024promptrpa}, enterprise automation~\citep{wornow2024automating}, financial question answering~\citep{zeng2023flowmind}, and data analysis~\citep{li2024autoflow}.
Despite relying on advanced LLMs (e.g., GPT-4), these approaches have often exhibited suboptimal performance, highlighting challenges faced by existing LLMs in workflow orchestration. 
While \citet{li2024autoflow} made an effort to fine-tune Mixtral-8$\times$7B~\citep{jiang2024mixtral}, it could only orchestrate sequential workflows with an average of $15.6$ actions, remaining insufficient for real-world requirements.
This work addresses a critical gap by proposing \WholeTitle framework to enhance the workflow orchestration capabilities of LLMs to meet real-world demands.

\paragraph{Tool Learning}
Workflow orchestration driven by LLMs frequently depends on external tools, such as APIs, to extend their operational capabilities. 
Recent studies have demonstrated that LLMs can effectively acquire and utilize external tools by learning from their documentation, thereby solving complex tasks that would otherwise be beyond the model's native capabilities~\citep{wu2023visual,schick2024toolformer,qin2023tool,DBLP:conf/iclr/QinLYZYLLCTQZHT24}. 
This integration enables LLMs to access real-time knowledge and perform specialized operations, particularly for executing intricate processes~\citep{yang2023chatgpt,nakano2021webgpt,qin2023webcpm,DBLP:journals/corr/abs-2407-19056,DBLP:conf/icml/GaoMZ00YCN23}.
To further enhance this capability, several efforts have introduced datasets specifically designed to fine-tune LLMs for tool interactions~\citep{zhuang2024toolqa,DBLP:conf/iclr/QinLYZYLLCTQZHT24,DBLP:journals/corr/abs-2402-01030}.
However, these datasets are often constrained to limited actions scale, thus limiting their effectiveness for managing complex, real-world workflows. 
Compared to tool learning scenarios, orchestrating workflows demands more sophisticated planning and reasoning that current LLMs have yet to fully realize.
In response to these limitations, we present \WholeTitle to significantly improve LLMs' capabilities in workflow orchestration.
Besides, \citet{shen2024shortcutsbench} also used Apple's Shortcuts but aimed to assess LLMs' tool utilization ability. 
%
%
%
In contrast, we emphasize a different scenario, workflow orchestration and aim to enhance the workflow orchestration ability rather than evaluation alone.

\begin{figure}[!t]
    \centering
    \includegraphics[width=0.9\linewidth]{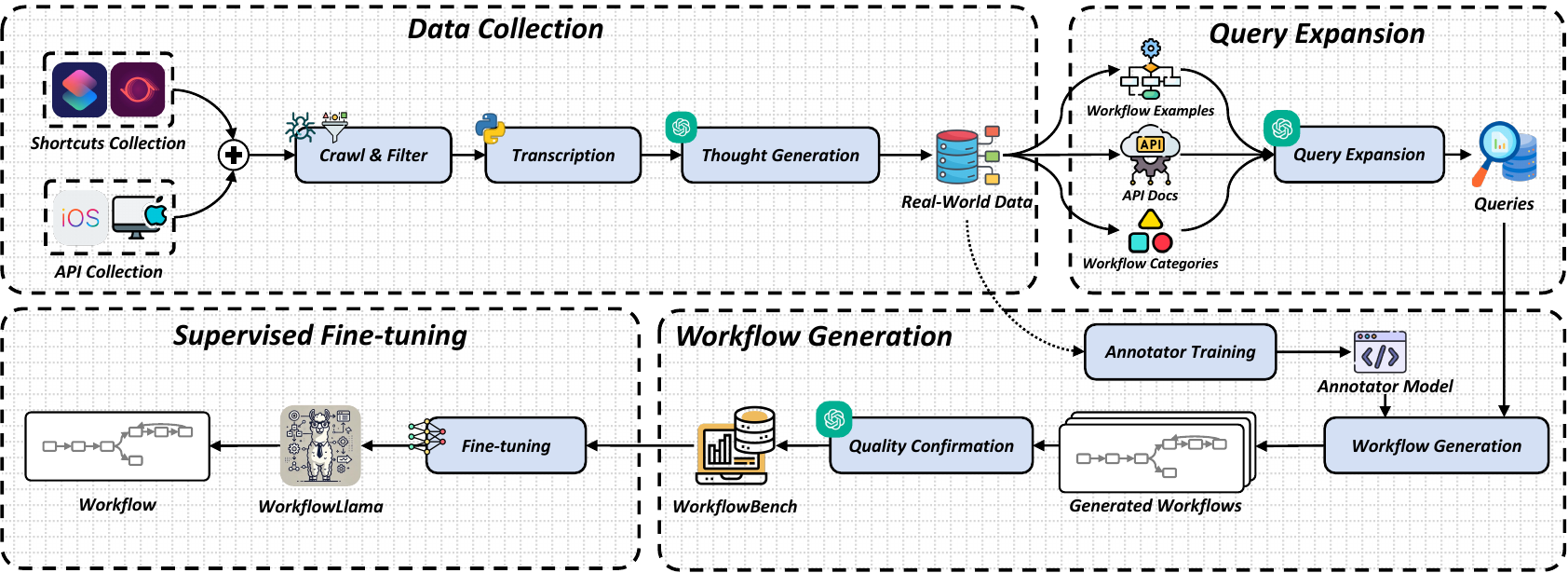}
    \caption{Illustration of our \WholeTitle which contains three phases to construct \BenchTitle, followed by the supervised fine-tuning phase to derive \ModelName.}
    \label{fig:dataset-construction}
\end{figure}

\section{\WholeTitle}

As \Cref{fig:dataset-construction} shows, \WholeTitle introduces a data-centric framework to enhance the capability of LLMs in workflow orchestration by constructing a high-quality supervised fine-tuning dataset \BenchTitle. 
In this section, we outline the dataset construction process, which is carried out in three distinct phases: Data Collection, Query Expansion, and Workflow Generation.

\subsection{Data Collection}
\label{sec:data-collection}

We first give the introduction to Apple Shortcuts and RoutineHub, and describe how we crawl and filter to get high-quality data.
We then convert the shortcuts into Python-style workflow code. 
Inspired by Chain-of-Thought~\citep{wei2022chain, chen2023program}, we prompt ChatGPT to generate hierarchical thoughts, including comments, task plans, and task queries, progressing from fine-grained to coarse-grained details for each shortcut.

\paragraph{Apple Shortcuts and RoutineHub}
Apple Shortcuts, as a representative application of RPA, is developed by Apple Inc. 
This tool facilitates the automation of a series of actions, enabling users to efficiently perform a diverse range of tasks.
The actions within Shortcuts are APIs provided by both built-in Apple applications, such as \textit{Safari}, and third-party applications like \textit{OpenAI}. Each application may provide multiple actions. 
For instance, \textit{OpenAI} provides APIs that facilitate voice conversations and text interactions with ChatGPT.
Through a simple drag-and-drop interface, users can construct complex workflows, such as navigating to the nearest coffee shop or downloading watermark-free images from TikTok.

RoutineHub\footnote{https://routinehub.co/} is a prominent community for sharing shortcuts, with a collection of thousands of shortcuts across both iOS and macOS platforms.
All shortcuts on RoutineHub are categorized into \CategoryNum workflow categories (e.g., Business, Health \& Fitness, Productivity, etc).
RoutineHub records the metadata of each shortcut (e.g., title, description, iCloud URL), providing valuable information.

\paragraph{Crawling and Filtering}
For each shortcut, we crawl the title, developer-provided description, and iCloud URL linked to Apple.
As RoutineHub does not provide the source code for these shortcuts, we further crawl it from their iCloud URLs.
Besides, we merge shortcuts collected by ShortcutsBench~\citep{shen2024shortcutsbench}, sourced from platforms like ShareShortcuts\footnote{https://shareshortcuts.com} and MacStories\footnote{https://www.macstories.net/shortcuts}, to further expand the scale of our dataset.
However, the source code of these shortcuts lacks detailed information about the involved actions, such as API metadata.
Inspired by ShortcutsBench~\citep{shen2024shortcutsbench}, we extract action information from macOS's built-in definition files and third-party application interface definition files. 
For each API, we record its name, description, parameter names, parameter types, default values, return value types, and return value name, which provides a valuable resource for LLMs to  efficiently interpret and utilize these APIs, even in zero-shot scenarios.

To ensure compatibility between the crawled shortcuts and the action interfaces, we implement a stringent filtering mechanism to verify that all API calls are executed correctly.
During this process, we identify that some shortcuts contain non-interpretable binary sequences as API parameters, potentially disrupting the training process of language models. 
To maintain data quality, we remove these samples from the dataset.  As a result, we curate a final set of \SeedNum high-quality shortcuts, ensuring the reliability of the dataset for subsequent data expansion and model training.

\paragraph{Shortcuts Transcription}
The original shortcut source codes are written in property lists format~\citep{Hummert2022}, which sequentially encodes logical constructs like branches and loops. This encoding is notably different from the types of data commonly used in the pre-training of LLMs.
To address this gap, we convert the shortcuts into abstract syntax trees (ASTs), apply pre-order traversal to transform them into Python code, with further algorithmic details provided in Appendix \ref{Transcribing}.
Furthermore, the original shortcuts use hexadecimal strings as variable names, leading to reduced semantic clarity. 
To improve interpretability, we use ChatGPT to automatically reassign these variables with more contextually meaningful names, thereby enhancing the overall readability and utility of the code for further language model training. 
A typical comparison between property lists and Python code can be found in Appendix \ref{sec::case_shortcuts}.

\paragraph{Thought Generation}
To provide informative guidance for LLMs in orchestrating workflows, we design a three-level thought hierarchy from fine-grained to coarse-grained:
(1) \textbf{Low-level comments} are intended to clarify the purpose of each action within the workflow.
(2) \textbf{Median-level plans} represent an abstraction over a sequence of actions, outlining the collective goal of these steps.
(3) \textbf{High-level queries} reflect the user's requirements, specifying the intended outcome without prescribing specific methods to achieve it.
These three levels of thought are generated through a bottom-up approach. 
Specifically, given the transcribed workflow $w$, let the set of actions in the workflow $w$ be denoted as $\mathcal{A}$, where each action $a_i \in \mathcal{A}$ corresponds to a function calling in the Python code. 
For each action $a_i$, we generate a corresponding comment $c_i$ by prompting ChatGPT.
Subsequently, given the action set $\mathcal{A} = \{ a_i\}$ and comments $\mathcal{C} = \{ c_i\}$ of workflow $w$, we prompt ChatGPT to generate the corresponding task plan $\mathcal{P}$.
We combine the task plan $\mathcal{P}$, the comments $\mathcal{C}$, and the action set $\mathcal{A}$ of the workflow $w$ to generate the high-level task query $\mathcal{Q}$.
This bottom-up manner is analogous to the summarization task, effectively ensuring content reliability and minimizing the risk of hallucination.

\begin{figure}[!t]
    \centering
    \includegraphics[width=0.9\linewidth]{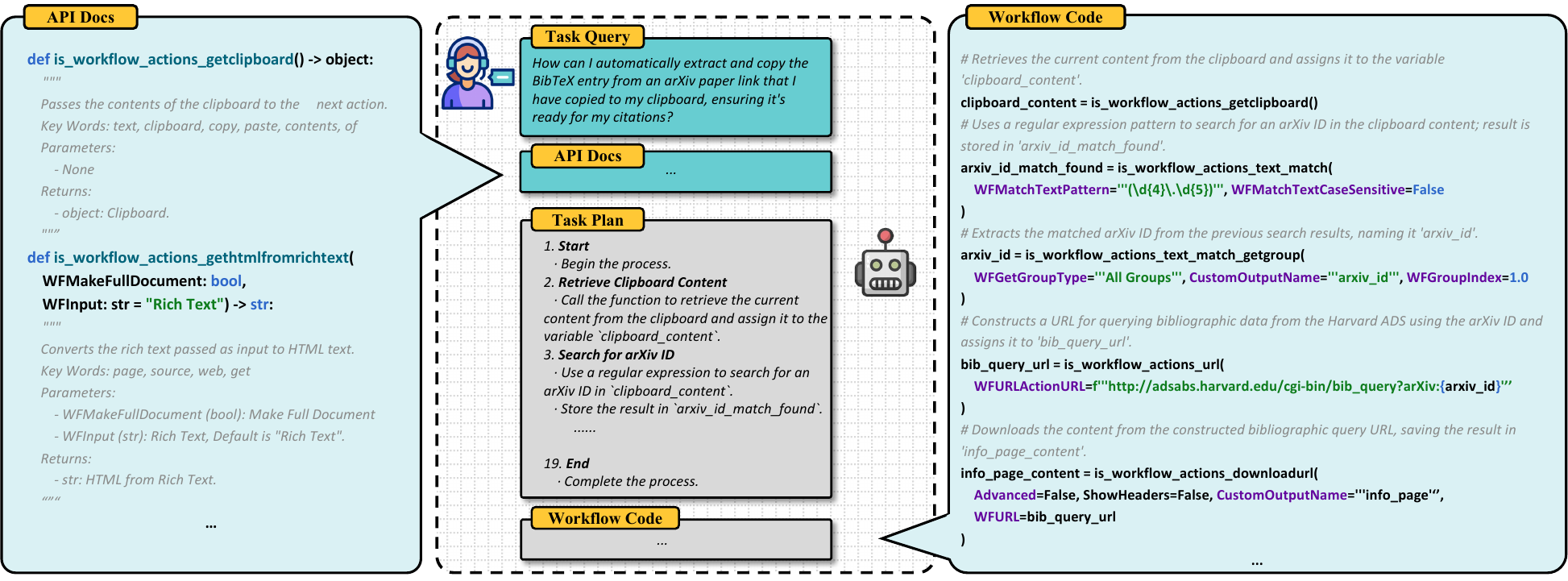}
    \caption{Illustration of data field composition in \BenchTitle comprising \textit{Task Query}, \textit{API documentations}, \textit{Task Plan}, and 
\textit{Workflow code} with \textit{Comments}.}
    \label{fig:train}
\end{figure}

Finally, as \Cref{fig:train} shows, each workflow $w$ is represented as: $w = \{ \mathcal{Q}, \mathcal{D}, \mathcal{P}, \mathcal{A} \}$, 
where the workflow $w$ consists of the task query $\mathcal{Q}$, action documentation $\mathcal{D}$ for all involved actions, the task plan $\mathcal{P}$, and all actions represented as annotated Python code $\mathcal{A}$. 
A detailed example can be found in Appendix \ref{case_workflow}.

\subsection{Query Expansion}
\label{sec:query-expansion}
After performing a comprehensive statistical analysis on the collected data, we find that the data exhibits significant complexity, with an average of 70.4 actions and 12 branches, surpassing the complexity of existing workflow-related benchmarks. However, the diversity of the data is relatively low. Specifically, 40.3\% of the workflows fall under the \texttt{Utilities} category, and over 99\% of the APIs used are Apple's built-in APIs (i.e., those classified as \texttt{is\_workflow\_actions} APP). 

Therefore, we intend to expand the dataset by focusing on two key aspects:
(1) \textbf{Diversity}: making up for the lack of diversity in real data and covering a broad range of APIs and workflow categories to enhance the model’s utility and robustness; (2) \textbf{Complexity}: matching the action scale and logical complexity of the real-world data to ensure that they can effectively represent real-world problems and orchestrate nodes accordingly.
To this end, we sample APIs from diverse applications and multiple workflows with representative logical structures (e.g., whether they contain branches or loops) to synthesize additional queries.

To ensure that the number of APIs in the synthesized dataset aligns with real-world usage, we sample \(n\) APIs based on real-world distributions. Approximately \(\lfloor n/2 \rfloor\) are drawn from Apple's built-in API set (e.g., \textit{openurl} or \textit{sendemail}), with the remainder from third-party applications (e.g., \textit{OpenAI}). The total number of built-in and external APIs is thus \(n\).

To ensure that the sampled APIs can interact coherently, we do not sample directly from the entire API set. 
Instead, we first randomly select 1-5 applications and then choose all APIs from these selected applications. 
This method ensures that the selected APIs are functionally compatible and capable of representing real-world workflows.

The prompt used for ChatGPT to synthesize queries consists of four components: (1) a general prompt to describe the task query generation task, (2) documentations for the sampled APIs, (3) in-context examples from the collected data for reference, and (4) the workflow category to which the query belongs to. 
By controlling the workflow category and in-context examples, we can ensure the diversity and complexity of the generated data. As seen from Figure \ref{fig:dist-cmp}, the synthesized query has a more balanced category distribution and uses more third-party APIs. Although most of the APIs used are still built-in APIs, this is reasonable considering that they carry necessary operations.

\subsection{Workflow Generation}
\label{sec:workflow-generation}

\begin{figure}[!t]
\centering
\includegraphics[width=1.0\linewidth]{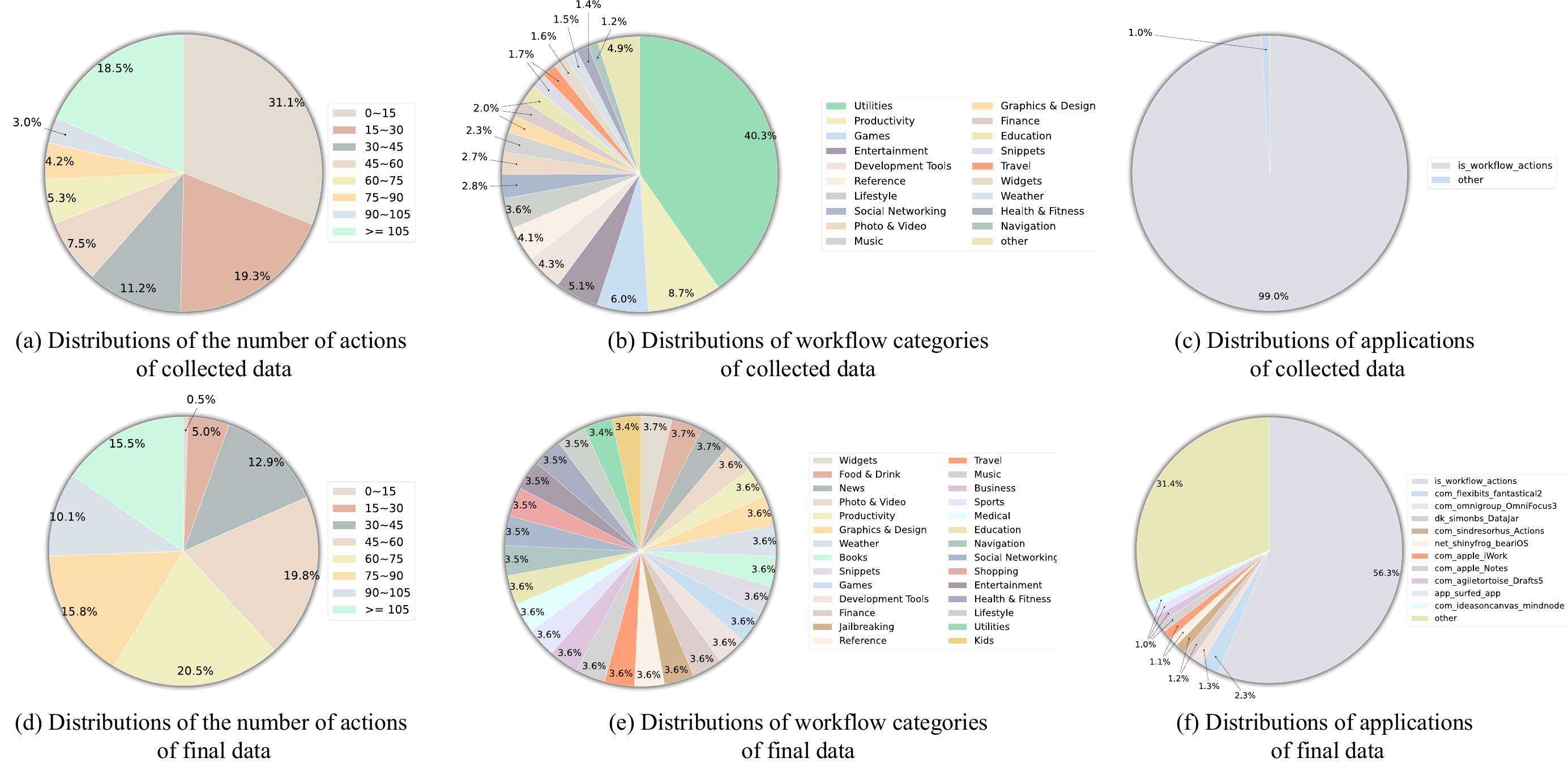}
\caption{
Comparison of the distributions across workflow categories, APPs, and action counts in the collected data and the final dataset. The upper section shows the original data collected from Apple Shortcuts and RoutineHub, while the lower section presents the expanded dataset distributions.
}
\label{fig:dist-cmp}
\end{figure}

\begin{wraptable}{r}{0.5\linewidth}
    \centering
    \footnotesize
    \renewcommand{\arraystretch}{1.1} 
    \vspace{-13pt}
    \caption{Detailed statistics of \BenchTitle. \textit{Seed.} refers to the collected data from Shortcuts. \textit{Train.} and \textit{Test.} refers to the training set and the test set of \BenchTitle respectively.}
    \begin{tabular}{l|r|r|r}
    \toprule
    Statistics & \multicolumn{1}{r|}{Seed.} & \multicolumn{1}{r|}{Train.} & \multicolumn{1}{r}{Test.} \\
    \midrule
    Num. of Instances & 14,771 & 105,573 & 1,190 \\
    Num. of APPs & 71 & 83 & 31 \\
    Num. of APIs & 584 & 1,503 & 324 \\
    Num. of Categories & 28 & 28 & 28 \\
    Avg. Action & 70.4 & 78.5 & 41.7 \\
    Avg. IF & 12.0 & 7.4 & 7.9 \\
    Avg. LOOP & 0.7 & 0.5 & 0.5 \\
    Avg. Nested Depth & 2.6 & 2.7 & 2.1 \\
    \bottomrule
    \end{tabular}
    \label{exp:stat-of-bench}
\end{wraptable}
To annotate the corresponding workflows of the synthesized queries effectively, we train an annotator model based on the collected shortcuts data to support more diverse applications and categories, while ensuring consistency with the real-world data as much as possible.

\paragraph{Annotator Training}
First, we construct the supervised fine-tuning~(SFT) dataset based on the collected human-labeled shortcuts. 
Specifically, each workflow data point comprises a query $\mathcal{Q}$, the corresponding action documentation $\mathcal{D}$, the task plan $\mathcal{P}$, and the workflow represented as annotated Python code $\mathcal{A}_{\text{commented}}$. 
During the SFT process, as shown in Figure \ref{fig:train},  we take the query $\mathcal{Q}$, the corresponding action documentation $\mathcal{D}$ as the input to guide the model to generate a task plan $\mathcal{P}$, followed by the step-by-step generation of the current thought (i.e., the comment $c_i$) and the corresponding action $a_i$, which includes the action name and its associated parameters.
We use the trained annotator to generate workflows $\mathcal{A'}$ from synthesized queries.

\paragraph{Quality Confirmation}
%
Due to the limited accuracy of the annotator model, the generated workflows may contain errors to some extent.
For example, we identify issues in $\mathcal{A'}$ (e.g., extraneous branches not relevant to the query and incorrect function call formats). 
To enhance the overall quality, we prompt ChatGPT with in-context samples to refine both $\mathcal{A'}_{\text{commented}}$ and $\mathcal{P'}$, ensuring that the workflow accurately addresses the query.
Then, we use rule-based filtering to remove workflows with fundamental errors. 
Specifically, we remove samples that don't incorporate code, don't utilize the given APIs, or violate parameter constraints associated with those APIs.
%

Finally, we derive a synthesized dataset of $91,992$ instances, which is combined with the initially collected data to form the final \BenchTitle.
It contains \TotalNum instances with \APINum APIs across \APPNum applications, which are used to train \ModelName.
The statistics of \BenchTitle are listed in \Cref{exp:stat-of-bench} and the distribution comparisons of workflow categories, APPs, and the number of actions between the collected data and final data are demonstrated in \Cref{fig:dist-cmp}.
From the statistical results, we can see that the synthetic data maintains complexity while expanding diversity.

\section{Experiments}

\subsection{Experimental Setup}

\textbf{Training Details\quad} We fine-tune the annotator and \ModelName on LLaMA-3.1-8B~\citep{dubey2024llama} for 3 epochs using the AdamW optimizer~\citep{DBLP:conf/iclr/LoshchilovH19}. A linear learning rate scheduler is used with a peak learning rate of $2 \times 10^{-5}$ and a warm-up ratio of $0.1$. Each mini-batch contains $32$ examples, and the maximum sequence length is set as $8,192$ tokens.

\textbf{Baselines\quad} 
To provide a comprehensive comparison, we select several representative LLMs as baselines for our experiments. These baselines include proprietary models such as GPT-4o-mini and GPT-4o, as well as open-source models like Qwen2-7B~\citep{qwen2}, Llama-3.1-8B, and Llama-3.1-70B~\citep{dubey2024llama}. 
Additionally, we apply in-context learning (ICL)~\citep{dong2022survey} with one random-sampled instance to these baselines to better adapt them for workflow orchestration.

\textbf{Metrics\quad } In the main experiments, we use both reference-code-based metrics and a model-based evaluation to comprehensively evaluate the quality of the generated workflows. 
For reference-based metrics, we apply \textbf{CodeBLEU}~\citep{ren2020codebleu} with four components: 
\begin{itemize}[topsep=1pt, partopsep=1pt, itemsep=-1pt, leftmargin=10pt]
    \item \textbf{BLEU} measures N-gram overlap for token-level similarity.
    \item \textbf{Weighted N-Gram Match} assigns higher weights to critical code tokens like keywords.
    \item \textbf{Syntactic AST Match} compares the Abstract Syntax Trees (ASTs) to assess syntactic accuracy.
    \item \textbf{Semantic Data-Flow Match} evaluates logical correctness by comparing data-flow relationships between variables.
    
\end{itemize}
Together, these components provide a comprehensive evaluation of both syntactic and semantic aspects of the workflows. 
We follow \citet{ren2020codebleu}, setting the four components to 0.1, 0.1, 0.4, and 0.4, respectively, and calculate a weighted sum to obtain the CodeBLEU score.
For model-based evaluation, we elaborately prompt ChatGPT as the automatic evaluator to evaluate the \textbf{Pass Rate} of the generated workflows.

\subsection{Effectiveness of Evaluator}
To validate the reliability of the ChatGPT evaluator in terms of Pass Rate, we sample 30 instruction-response pairs (i.e., task queries and their corresponding workflow codes) for each model in Table~\ref{exp:main}, forming a human-evaluated dataset of $330$ instances (\(30 \times 11 = 330\)). First, we use GPT-4o-mini to label whether each instance could complete the given tasks only using the provided APIs. Then, human evaluators re-label the sampled data according to the same criteria. Ultimately, $268$ instances are labeled consistently by both the ChatGPT evaluator and human evaluators, achieving an agreement rate of $\textbf{81.2}\%$, demonstrating the reliability and effectiveness of the evaluator.

\begin{table}[!t]
\centering
\footnotesize 
\caption{Performance comparison of various models on the test set of \BenchTitle\ under the \textbf{unseen instructions (ID)} and \textbf{unseen APIs (OOD)} settings (\%).}
\begin{widetabular}{\textwidth}{lcccccccccccc}  
\toprule
\multirow{3}{*}{\textbf{Model}} & \multicolumn{10}{c}{\textbf{CodeBLEU}} & \multicolumn{2}{c}{\multirow{2}*{\textbf{Pass Rate}}} \\
\cmidrule{2-11}
  & \multicolumn{2}{c}{\underline{\textbf{BLEU}}} & \multicolumn{2}{c}{\underline{\textbf{Weighted N-Gram}}} & \multicolumn{2}{c}{\underline{\textbf{AST}}} & \multicolumn{2}{c}{\textbf{\underline{Data-Flow}}} & \multicolumn{2}{c}{\underline{\textbf{Overall}}} &  &  \\
  & \textbf{ID} & \textbf{OOD} & \textbf{ID} & \textbf{OOD} & \textbf{ID} & \textbf{OOD} & \textbf{ID} & \textbf{OOD} & \textbf{ID} & \textbf{OOD} & \textbf{ID} & \textbf{OOD} \\
\midrule
\multicolumn{13}{c}{\cellcolor{pink!30}\textbf{Proprietary Models}} \\ 
\midrule
GPT-4o-mini     & 0.4  & 0.4  & 1.5  & 1.6  & 29.5  & 29.5  & 37.0  & 36.3  & 26.8  & 26.5  & 54.8  & 47.5 \\
\quad \textit{w/ ICL}      & 0.5  & 0.5  & 1.7  & 1.8  & 35.3  & 34.4  & 35.1  & 34.2  & 28.3  & 27.7  & 66.0  & 57.7 \\

GPT-4o          & 0.5  & 0.4  & 1.8  & 1.7  & 33.5  & 31.8  & 37.3  & 36.9  & 28.5  & 27.7  & 56.6  & 47.5 \\
\quad \textit{w/ ICL}      & 0.5  & 0.5  & 1.8  & 1.8  & 37.1  & 35.3  & 38.0  & 36.6  & 30.2  & 30.0  & 67.5  & 57.6 \\
\midrule
\multicolumn{13}{c}{\cellcolor{blue!15}\textbf{Open-Source Models}} \\ 
\midrule
Qwen2-7B       & 0.4  & 0.4  & 1.2  & 1.3  & 27.2  & 27.7  & 33.2  & 33.1  & 24.4  & 24.5  & 25.6  & 22.6 \\
\quad \textit{w/ ICL}      & 0.5  & 0.5  & 1.2  & 1.3  & 30.2  & 29.8  & 32.4  & 32.9  & 25.2  & 25.3  & 28.2  & 26.4 \\
Llama-3.1-8B   & 0.6  & 0.7  & 1.2  & 1.4  & 31.0  & 29.6  & 30.0  & 30.8  & 24.6  & 24.3  & 33.0  & 24.5 \\
\quad \textit{w/ ICL}      & 0.7  & 0.7 & 1.3  & 1.4  & 34.0  & 32.4  & 32.6  & 32.4  & 25.3  & 25.2  & 40.2  & 32.7 \\
Llama-3.1-70B  & 0.4    & 0.4    & 1.4    & 1.5    & 29.9     & 30.0     & 37.8     & 37.6    & 27.3     & 27.2    & 55.4  & 42.3 \\
\quad \textit{w/ ICL}      & 0.4  & 0.4  & 1.6  & 1.5  & 34.1  & 32.9  & \textbf{39.1}  & \textbf{38.4}  & 29.5  & 28.7  & 67.6  & 61.4   \\
\midrule
\ModelName (8B)  & \textbf{9.4} & \textbf{7.0} & \textbf{11.09} & \textbf{8.3} & \textbf{55.1} & \textbf{48.8} & 38.0 & 35.3 & \textbf{39.3} & \textbf{35.1} & \textbf{76.9}  & \textbf{70.4} \\
\bottomrule
\end{widetabular}
\label{exp:main}
\end{table}

\subsection{Main Experiments}
\textbf{Settings\quad}
The main experiments are conducted using the test set of \BenchTitle. Ideally, by scaling both the quantity and diversity of instructions and unique tools within the training data, \ModelName is expected to generalize to novel instructions and APIs that are not seen during training. 
This is particularly important because it enables users to define custom APIs and allows \ModelName to adapt based solely on the provided documentation.
To evaluate this capability, we assess \ModelName's generalization performance at two levels: (1) \textbf{Unseen Instructions}, considers an \textbf{In-Distribution (ID)} setting, which involves using the same set of APIs as those in the training data, and (2) \textbf{Unseen APIs}, considers an \textbf{Out-Of-Distribution (OOD)} setting, involving only 50 common APIs required to construct workflows and APIs that are absent from the training data.
Since \BenchTitle contains a comprehensive set of APIs, which poses a substantial challenge for LLMs in terms of API comprehension and selection, we provide the correct APIs directly as input. It allows us to focus on the workflow orchestration, bypassing the issue of API selection.

\textbf{Main Results\quad}
The results are placed in Table~\ref{exp:main}, from which we derive that:

\begin{enumerate}[topsep=1pt, partopsep=1pt, leftmargin=12pt, itemsep=-1pt] 
    \item Although multiple workflows can successfully complete a query, there is a positive correlation between the reference-free Pass Rate metric and the reference-based CodeBLEU metric. Given that the Pass Rate metric derived from ChatGPT aligns with human evaluations over 80\% of the time, CodeBLEU serves as a reliable proxy for evaluating workflow orchestration capabilities.

    \item All models demonstrate a certain capacity for workflow orchestration. This may stem from their inherent instruction-following and code-generation capabilities.
    We find that models like GPT-4o and Llama-3.1-70B, which perform better on generic tasks, also excel in workflow orchestration.
    In addition, prompting with in-context samples significantly enhances the models' performance.
    \item We find that scores on text overlap metrics such as BLEU and weighted N-gram are low for all models. Even the fine-tuned \ModelName only achieves $8.2\%$ and $9.7\%$ on these two metrics. 
    This is because the reference codes consist mainly of workflows with function names and arguments, and contain few Python-related keywords, making exact matching challenging.
    In contrast, models achieve better scores on syntactic AST match and semantic data-flow match.
        
    \item After fine-tuning, \ModelName\ shows a significant improvement in its ability to orchestrate actions. The performance of \ModelName even outperforms powerful closed-source models GPT-4o with ICL by a large margin. Specifically, \ModelName achieves a $\textbf{39.3}\%$ score on CodeBLEU and a $\textbf{76.9}\%$ Pass Rate under ID settings,  demonstrating the validity of our proposed \WholeTitle framework and \BenchTitle dataset.

    \item \ModelName demonstrates strong generalization capabilities. Even though it has not been trained on the same instructions or APIs, it still significantly outperforms the vanilla Llama-3.1 on all metrics, ahead of or close to the more powerful foundation models. 
    Notably, our method achieves $\textbf{35.1}\%$ in CodeBLEU and $\textbf{70.4}\%$ in Pass Rate, outperforming all strong baselines.
\end{enumerate}

\subsection{Analysis of Workflow Complexity}

\begin{figure}[!t]
    \centering
    \begin{subfigure}[b]{0.32\textwidth}
        \centering
        \includegraphics[width=\textwidth]{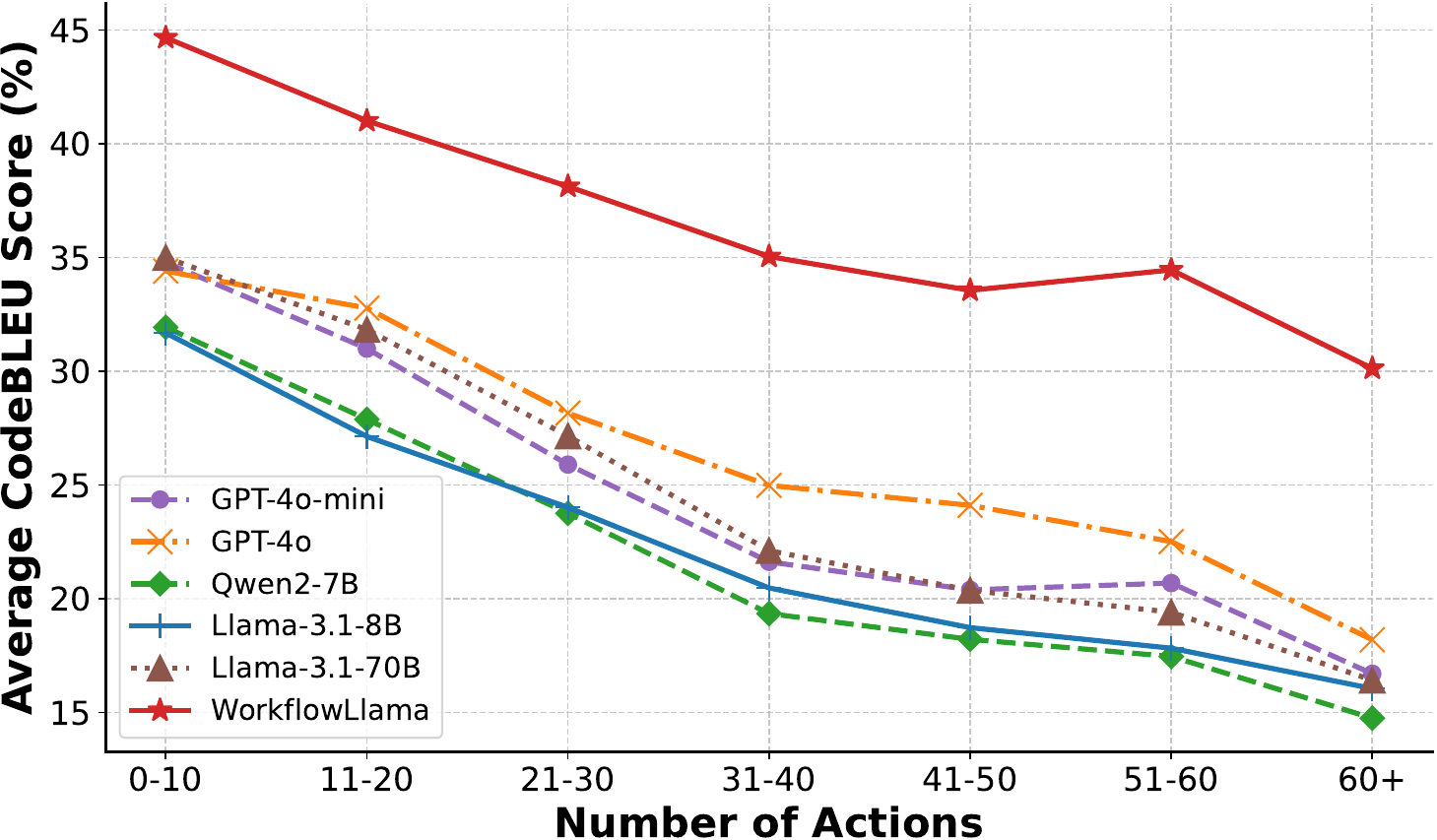}
    \end{subfigure}
    \hfill
    \begin{subfigure}[b]{0.32\textwidth}
        \centering
        \includegraphics[width=\textwidth]{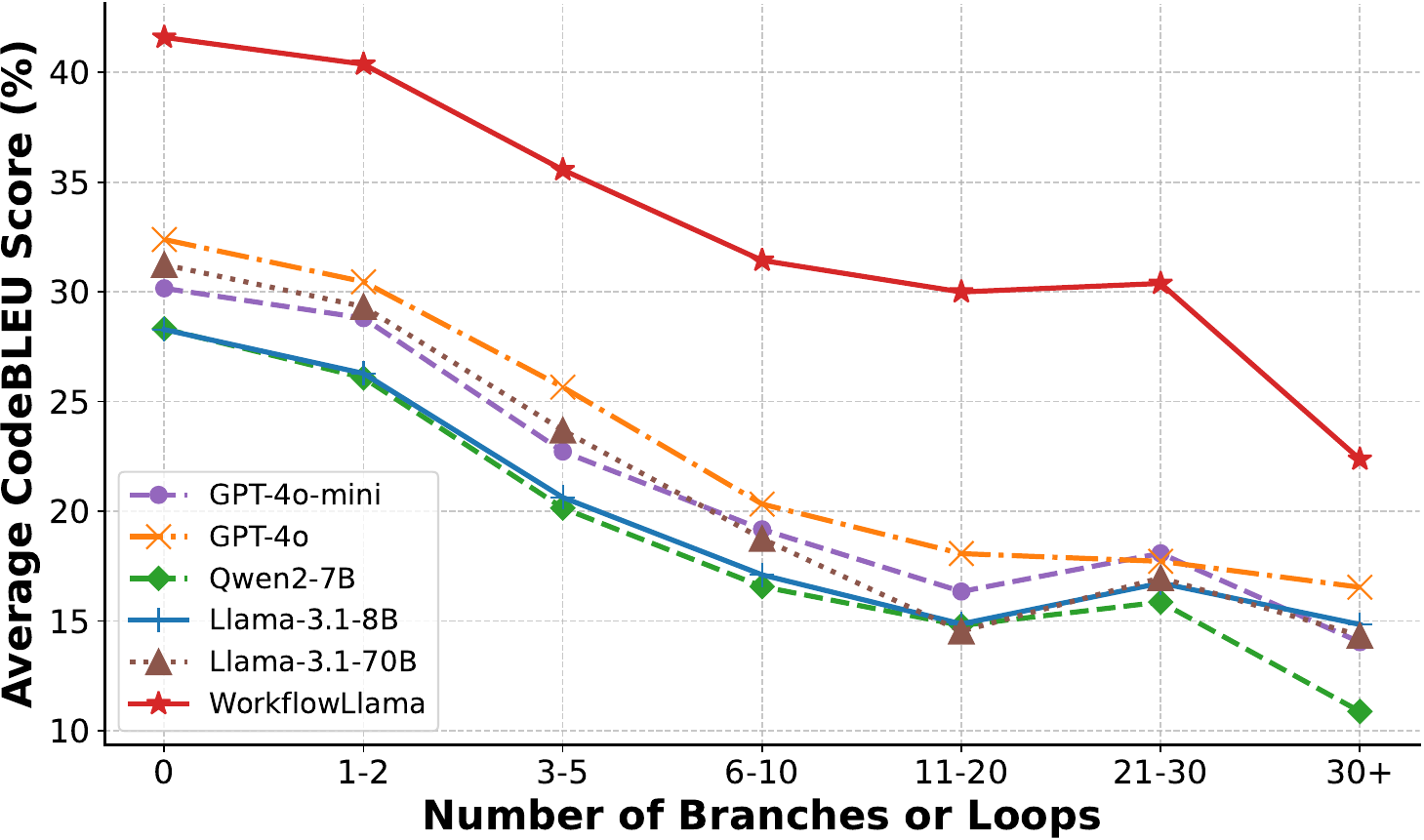}
    \end{subfigure}
    \hfill
    \begin{subfigure}[b]{0.32\textwidth}
        \centering
        \includegraphics[width=\textwidth]{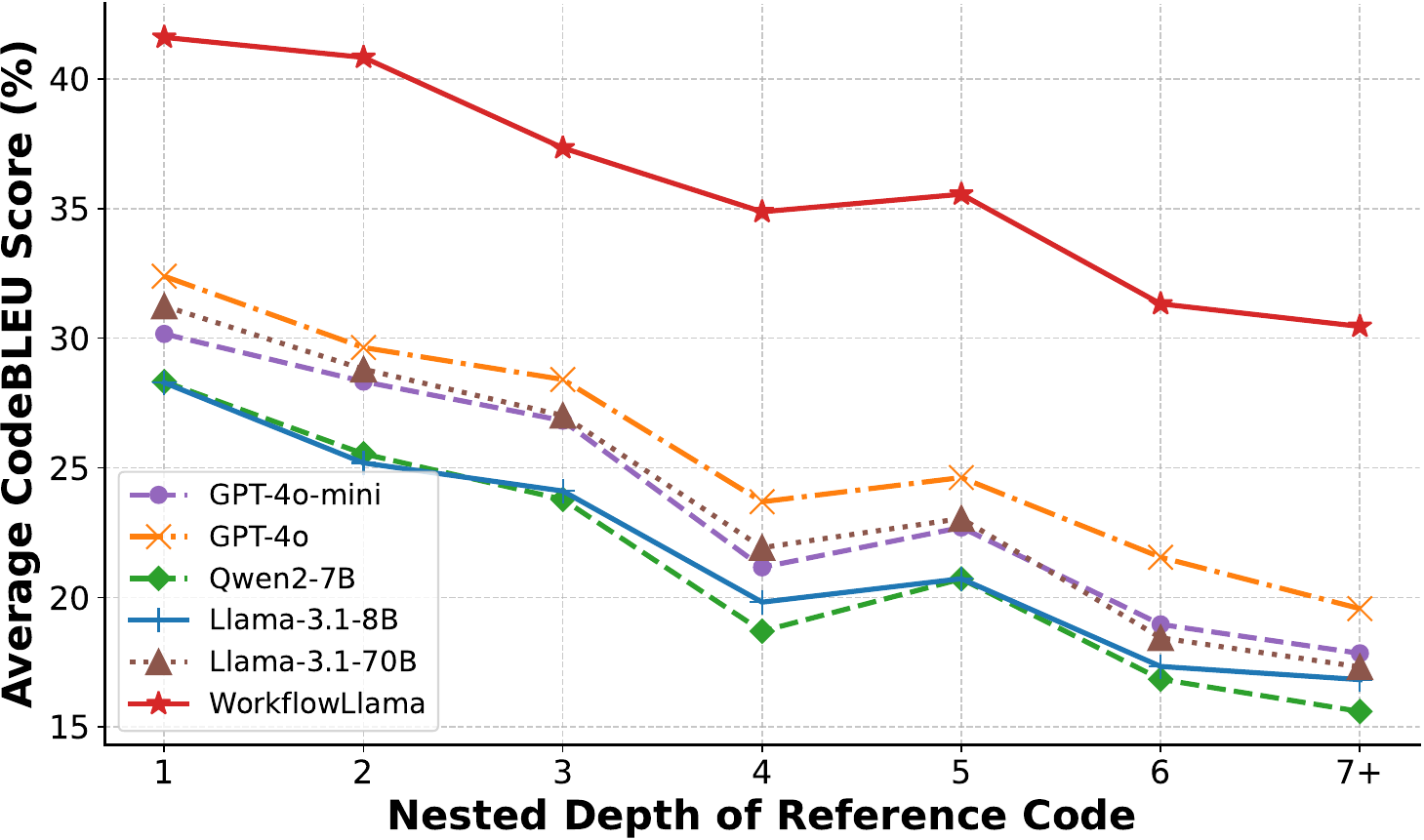}
    \end{subfigure}
    \caption{Performance comparisons based on the number of actions, the number of Branch \& Loop, and the nested depth of the reference code.}
\label{fig:analysis}
\end{figure}
To evaluate the models' ability to generate workflows of varying complexity, we break down the performance of CodeBLEU according to the total number of actions, the number of branches and loops, and the nested depth of the reference code. As shown in Figure~\ref{fig:analysis}, the performance of all models deteriorates as the number of actions or the logical complexity increases, indicating the challenge of orchestrating complex workflows.
However, across all levels of complexity, \ModelName significantly outperforms all other models. Moreover, the relative performance of \ModelName improves as the complexity of the workflow increases, which demonstrates fine-tuning with \BenchTitle significantly enhances the model’s ability to handle more complex workflows.

\subsection{Out-of-distribution Generalization to T-Eval \citep{chen2024t}} 

\begin{wraptable}{r}{0.5\textwidth} 
\centering
\vspace{-12pt}
\footnotesize
\setlength{\tabcolsep}{12pt} 
\caption{Comparisons of F1 scores on the \textbf{PLAN} task of T-Eval. (\textbf{Bold} denotes the best score among models of the same category.)}
\begin{tabular}{p{0.5\linewidth} p{0.15\linewidth}} 
\toprule
\textbf{Model} & \textbf{F1} \\ \midrule
\multicolumn{2}{c}{\cellcolor{pink!30}\textbf{Proprietary Models}} \\ 
\midrule
Claude2    & 84.9 \\
GPT-3.5    & 86.6 \\
GPT-4      & \textbf{86.7} \\
\midrule
\multicolumn{2}{c}{\cellcolor{blue!15}\textbf{Open-Source Models}} \\ 
\midrule
Qwen-7B      & 63.1 \\
Mistral-7B   & 64.9 \\
Llama-3.1-8B  & 68.2 \\
\midrule
Qwen-14B     & 69.7 \\
Llama-2-13B  & 65.1 \\
Vicuna-13B   & 54.0 \\
Baichuan2-13B& 52.1 \\
\midrule
WizardLM-70B & 42.7 \\
Llama-2-70B  & 63.1 \\
Qwen-72B     & 73.4 \\
\midrule
\ModelName (8B)   & \textbf{77.5} \\
\bottomrule
\end{tabular}

\label{tab:OOD}
\end{wraptable}
\textbf{Settings\quad} 
To further evaluate the generalization capability of \ModelName, we conduct experiments using an OOD benchmark, T-Eval, a widely-used benchmark to evaluate the multi-step decision-making capability of LLMs to utilize APIs. 
The original data format in T-Eval is based on JSON or strings, which differ significantly from the Python-based format employed in \BenchTitle. To ensure the evaluation metrics’ consistency between ours and the original paper, we convert \BenchTitle into JSON format while preserving the metadata of workflows and the specifics of queries. Subsequently, we retrain \ModelName on the transformed dataset.
We employ the \textbf{F1 Score} proposed in the original paper to measure the alignment with the reference API sequences.

\textbf{Results\quad} The results are shown in Table~\ref{tab:OOD}. As observed, \ModelName demonstrates strong OOD generalization performance on the T-Eval benchmark, despite being trained on different domains and tasks using different APIs. Notably, \ModelName significantly outperforms the vanilla Llama3.1-8B as well as larger open-source models like Llama-2-70B and Qwen-72B, highlighting that fine-tuning with \BenchTitle enhances the model's out-of-distribution planning ability.

\subsection{Ablation Study} 

\begin{table}[!t]
\centering
\footnotesize
\caption{Ablation study results of Natural Language Thoughts on Workflow Orchestration (\%).}
\begin{widetabular}{\textwidth}{lccccc}
\toprule
\multirow{2}*{\textbf{Model}} & \multicolumn{5}{c}{\textbf{CodeBLEU}} \\
\cmidrule{2-6}
               & \textbf{BLEU} & \textbf{Weighted N-Gram} & \textbf{AST} &  \textbf{Data-Flow} & \textbf{Overall} \\
\midrule
\ModelName     & 9.4           & 11.1                    & 55.1         & 38.0                & 39.3              \\
\ \quad\textit{w/o Task Plan} & 9.1           & 10.7                    & 53.9         & 36.6                & 38.2              \\
\ \quad\textit{w/o Comment}  & 9.1           & 10.8                    & 54.9         & 35.3                & 38.1              \\
\ \quad\textit{w/o Task Plan \& Comment}      & 8.8           & 10.2                    & 53.7         & 35.1                & 37.4              \\
\ \quad\textit{w/o Synthetic Data}      & 7.8           & 9.4                     & 53.5         & 35.4                & 37.3             \\
\bottomrule
\end{widetabular}
\label{exp:ablation}
\end{table}
\textbf{Settings\quad} 
To assess the efficacy of WorflowBench’s components, we conduct an ablation study under the settings of unseen instructions (i.e., the ID setting).

\textbf{Results\quad}
Table \ref{exp:ablation} presents the performance results when the model is trained under different conditions: without synthetic data, without the task plan $\mathcal{P}$, without action-level comments $\mathcal{C}$, and without both $\mathcal{C}$ and $\mathcal{P}$. The experimental results reveal two key findings.
\textbf{First}, the two types of natural language thoughts enhance the reasoning capabilities of the model. Removing either type of thought leads to a decline in CodeBLEU performance.
\textbf{Second}, training on large-scale synthetic data further improves performance, highlighting the effectiveness of the \BenchTitle expansion process.

\subsection{Case Study}

\begin{figure}[!t]
    \centering
    \includegraphics[width=1.0\linewidth]{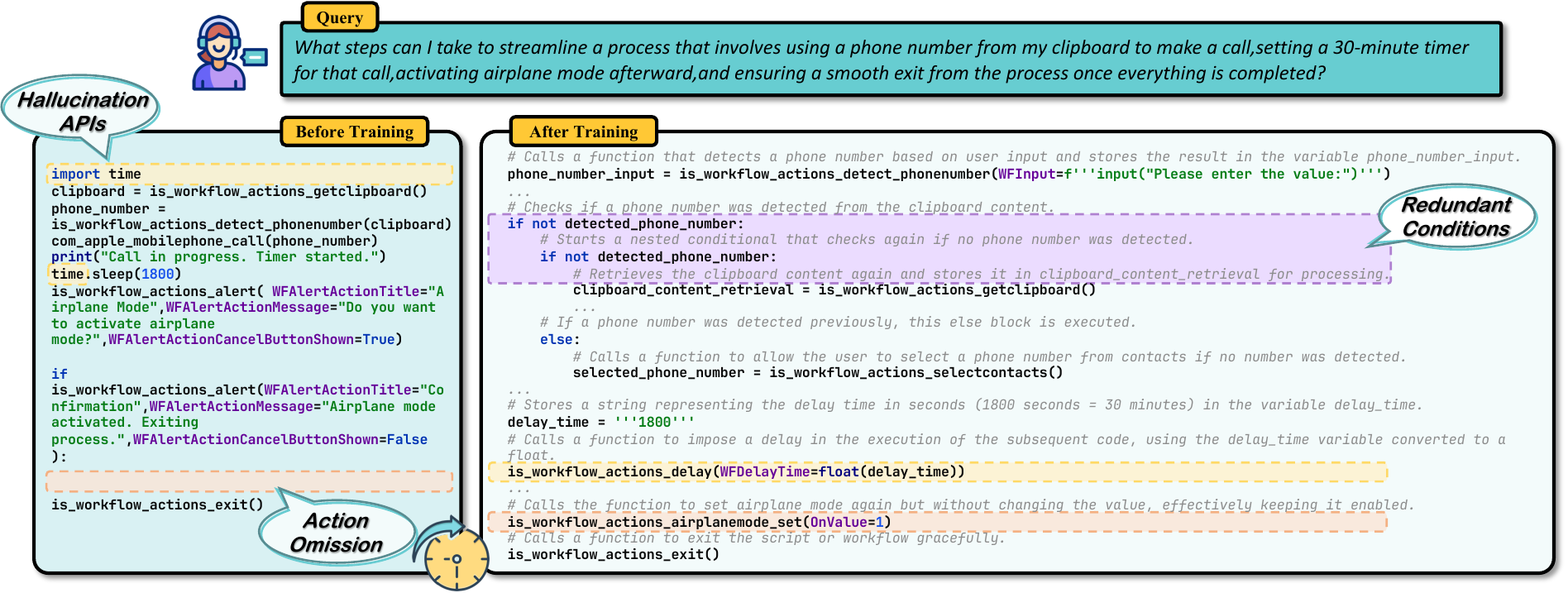}
    \caption{Case study of generated code between vanilla Llama-3.1-8B and \ModelName.}
    \label{fig:case_study}
\end{figure}
To further illustrate the effect of fine-tuning on \BenchTitle, we present a typical example in Figure \ref{fig:case_study}. In this case, the vanilla Llama-3.1 model exhibits two types of errors. 
\textbf{First}, the model does not adhere to the given instructions for workflow orchestration, using APIs outside the provided list, i.e., hallucination APIs. Specifically, it uses the \texttt{time.sleep()} function instead of \texttt{is\_workflow\_actions\_delay()} to set a timer. 
\textbf{Second}, due to its relatively weak workflow orchestration capabilities, the model fails to complete all user instructions. Specifically, it does not activate airplane mode using the \texttt{is\_workflow\_actions\_airplanemode\_set()} function.
Fine-tuning on \BenchTitle effectively alleviates these two issues. However, we observe that fine-tuning also introduces redundant actions. For instance, \ModelName repeats the parsing check of the clipboard's content. We will address this redundancy problem in future work.

\section{Conclusion}
In this paper, we present \WholeTitle to enhance the capability of large language models in workflow orchestration.
In \WholeTitle, \BenchTitle is constructed covering \TotalNum workflows with \APINum APIs across \APPNum applications through a three-phase pipeline.
By fine-tuning Llama-3.1-8B on \BenchTitle, we derive \ModelName which can achieve superior performance on the workflow orchestration task exceeding all comparable baselines including the most advanced OpenAI GPT-4o.
Moreover, we adapt our \ModelName on the T-Eval dataset and the experimental results reveal the generalization ability of our constructed \BenchTitle.
We believe that our constructed dataset has the potential to contribute to advancements in APA.

\bibliography{iclr2025_conference}
\bibliographystyle{iclr2025_conference}

\appendix

\section{Algorithm of Transcribing Shortcuts}\label{Transcribing}
\begin{algorithm}[H]
\DontPrintSemicolon
\KwData{Shortcut file to be transcribed}
\KwResult{Abstract syntax tree of the actions}

Initialize an empty tree with a root node and set \texttt{current\_node} to root\;

\ForEach{action \textbf{in} action list}{
    Determine \texttt{action\_type} and \texttt{mode} from action\;
    \uIf{\texttt{action\_type} is \textbf{Conditional}}{
        \textbf{HandleConditional}(\texttt{mode}, action)\;
    }
    \uElseIf{\texttt{action\_type} is \textbf{RepeatEach}}{
        \textbf{HandleLoop}(\texttt{mode}, action)\;
    }
    \uElseIf{\texttt{action\_type} is \textbf{RepeatCount}}{
        \textbf{HandleLoop}(\texttt{mode}, action)\;
    }
    \uElseIf{\texttt{action\_type} is \textbf{ChooseFromMenu}}{
        \textbf{HandleMatchCase}(\texttt{mode}, action)\;
    }
    \Else{
        \textbf{HandleDefault}(action)\;
    }
}

\SetKwProg{Fn}{Function}{:}{}
\Fn{\textbf{HandleConditional}(\texttt{mode}, action)}{
    \uIf{\texttt{mode} == 0 (\textbf{start if})}{
        \textbf{AddNode}(action)\;
        Set \texttt{current\_node} to new node\;
    }
    \uElseIf{\texttt{mode} == 1 (\textbf{else})}{
        Move \texttt{current\_node} to parent node\;
        \textbf{AddNode}(action)\;
        Set \texttt{current\_node} to new node\;
    }
    \uElseIf{\texttt{mode} == 2 (\textbf{end if})}{
        Move \texttt{current\_node} to parent node\;
    }
}

\Fn{\textbf{HandleLoop}(\texttt{mode}, action)}{
    \uIf{\texttt{mode} == 0 (\textbf{start loop})}{
        \textbf{AddNode}(action)\;
        Set \texttt{current\_node} to new node\;
    }
    \uElseIf{\texttt{mode} == 2 (\textbf{end loop})}{
        Move \texttt{current\_node} to parent node\;
    }
}

\Fn{\textbf{HandleMatchCase}(\texttt{mode}, action)}{
    \uIf{\texttt{mode} == 0 (\textbf{start match})}{
        \textbf{AddNode}(action)\;
        Set \texttt{current\_node} to new node\;
    }
    \uElseIf{\texttt{mode} == 1 (\textbf{start case})}{
        \uIf{\texttt{current\_node} is match node}{
            \textbf{AddNode}(action)\;
            Set \texttt{current\_node} to new node\;
        }
        \Else{
            Move \texttt{current\_node} to parent match node\;
            \textbf{AddNode}(action)\;
            Set \texttt{current\_node} to new node\;
        }
    }
    \uElseIf{\texttt{mode} == 2 (\textbf{end match})}{
        Move \texttt{current\_node} to parent node\;
    }
}

\Fn{\textbf{HandleDefault}(action)}{
    \textbf{AddNode}(action)\;
}

\Fn{\textbf{AddNode}(action)}{
    Create new node with \texttt{action}\;
    Append new node to \texttt{current\_node.children}\;
    Set parent of new node to \texttt{current\_node}\;
}

\caption{Recursive Parsing of Property List to Construct Abstract Syntax Tree}
\end{algorithm}

\section{Prompt Design}

\subsection{Workflow Orchestration Prompt}
\begin{lstlisting}[basicstyle=\ttfamily, breaklines=true]
You are a very helpful AI assistant who can write corresponding Python main code based on user's query and usable Python function interface.

Please generate python main code based on the following query :
 {query}
You can start by using natural language to plan your tool call strategy, and then generate the complete code. For example, `Thought:
<tool call strategy>

Code:
```python
<main code>
````.
Note that your output should always include `Code:
```python
<main code>
````, formatted accordingly.
Here are some useful function interface you may use:
 {apis_docs}
\end{lstlisting}

\subsection{Evaluator Prompt}
\begin{lstlisting}[basicstyle=\ttfamily, breaklines=true]
You are a kindly code reviewer, I will provide you with a query, a list of allowed apis and a piece of code to be reviewed, you help me to check if the code to be reviewed is compliant with our specifications.
The requirements are as follows:
1. You **should return True even if the code implements additional functionality not required in the query**, as long as it roughly implements the requirements in the query.
2. We don't impose any requirements on code readability or naming conventions. You **should return True as long as the reviewed code doesn't use disallowed functions and reasonably accomplishes what is asked in the query in general terms**. There's no need to get strictly hung up on the details.
3. Return False if the code fails to fulfill the requirement in the query. e.g. if it is proposed in the query to turn down the battery level of the phone and the brightness of the screen, it is a failure to fulfill only any one of the functions.
4. Built-in python syntax such as `if`, `loop`, `input()`, and `print()` are allowed.  Return False if the code uses **any external functions or apis** not in allowed apis list and not a built-in function such as input(), print(). For example, if I provide the is_workflow_openurl function, this should be used. Any use of any other library like requests etc. is a False.
query:{query}
list of allowed apis: {apis}
code to review: {code}

Your answer: [True or False with interpretation]
\end{lstlisting}

\subsection{Comment Generation Prompt}
\begin{lstlisting}[basicstyle=\ttfamily, breaklines=true]
A Shortcut is a sequence of actons, where each action is an API call, to execute user-provided queries.
As a user-friendly and patient assistant, your task is to provide a set of description of each line of the code scrippet. To save time, I have retrieved all the lines exclusive of blank lines of the code snippet and listed as a dictionary below the code.

Your answer should be in the json format as follows:
```json
{
    "line x": "<description-of-line-x>",
    "line x+1": "<description-of-line-x+1>",
    "...": "...",
    "line x+n": "<description-of-line-x+n>"
}```

The code is :
{code}
The lines are {lines}
\end{lstlisting}

\subsection{Task Plan Generation Prompt}
\begin{lstlisting}[basicstyle=\ttfamily, breaklines=true]
Based on this line by line description of the code, generate a flowchart of a workflow by natural language.
This is the code:
{code}
\end{lstlisting}

\subsection{Task Query Generation Prompt}
\begin{lstlisting}[basicstyle=\ttfamily, breaklines=true]
As a helpful assistant, please help me craft a query. This query, formatted as a question, should describe the task a user wants to complete and adhere to the following criteria:         
1. One of the solution to the task described in the query could be the python code below. 
2. It should be close to real-world problems or requests.
3. It should include major parts of the code.
4. The query should not specify python.

For example, the code is:
{ICL_code}
And the expected output query should be similar to:
{ICL_query}

Please craft a query based on the examples and the following code:
{code}
\end{lstlisting}

\subsection{Query Expansion Prompt}
\begin{lstlisting}[basicstyle=\ttfamily, breaklines=true]
You are exceptionally skilled at crafting real-world user queries given some apis. Here are examples:{examples}. Please gain inspiration from the following api docs to create a high-quality realworld query.
Api docs for inspiration:
```python
{apis_string}
```
Please refer to the above examples and craft a new one!
Requirements: API name is strictly prohibited from appearing in the generated query. Each query should be complicated enough and can be solved using all apis above. The query **should be centered around {category} theme** and should not be spread out into unrelated pieces.
\end{lstlisting}

\subsection{Quality Confirmation Prompt}
\begin{lstlisting}[basicstyle=\ttfamily, breaklines=true]
You are exceptionally skilled at polishing tool calling plan (i.e., thought) and python code given a task. 

Given task:
{query}


Old tool calling plan:
{thought}

 Old code:
{code}

 Used API doc:
{apis}

Here are examples for you to refer:{ICL_context}.
Please make sure the code is logically correct and operational. 

Requirements:
[1] Ensure that both plan and code respond correctly to the task and that code calls match the plan, which you can do by tweaking, embellishing, and modifying both plan and code.
Plan does not have to be one-to-one correspondence of code; plan can be abbreviated.
[2] Please ensure that the code conforms to python syntax. Ensure that all python code is complete and runnable. You can add code when necessary.
[3] Every line of code should be preceded by a comment marked with a "#". When modifying the code, please modify the in-line comments accordingly.
[4] Ensure that all function parameter calls are correct and you can change the code in case of errors.
[5] Thought and code should be as concise while keeping the meaning intact.
[6] If there are cases including invalid binary code, replace them with reasonable text, delete them, or replace them with a reading operation on a file (especially when the binary code is an encoded image).
Respond strictly with JSON.

\end{lstlisting}

\subsection{Variable Rename Prompt}
\begin{lstlisting}[basicstyle=\ttfamily, breaklines=true]
You are a helpful assistant for renaming variable names in a code snippet. 
The following code snippet is a part of a program, and variables are named in format 'variablex_'. 
Your task is to rename these variables so that they conform to the programming specification and have some semantic meaning, which can be infered by relative function calls
And your output should only be a dictionary containing the old name-new name key value pair
The definition of some functions are not included, and you shouldn't modify them. 
Following the code, there's a dictionary that contains short description of the uuid-named variable. And you can take it as reference. 
Note that while the description might be the same, but the actual meaning is different across different variables. So you should not just copy the short description. Instead you'd better conprehensively consider the description, names of called functions, and the general logic.
The code is as follows:
{code}
The dictionary is as follows:
{description}
To save time, I have retrieved all the variables that requires to be renamed:
{variables}
\end{lstlisting}

\section{Case Study of Shortcuts} \label{sec::case_shortcuts}
We provide a real-world shortcut example, which includes the following three presentation forms: the rwa property list configuration file, the Python code after transcription and variable renaming, and the visual interface on MacOS.

The raw property list configuration file is presented below. For the sake of brevity, we have omitted the middle portion containing the actions.
\small 
\begin{verbatim}

{
  "WFWorkflowClientVersion": "754",
  "WFWorkflowClientRelease": "2.1.2",
  "WFWorkflowMinimumClientVersion": 411,
  "WFWorkflowIcon": {
    "WFWorkflowIconStartColor": 4274264319,
    "WFWorkflowIconImageData": "b''",
    "WFWorkflowIconGlyphNumber": 59672
  },
  "WFWorkflowImportQuestions": [],
  "WFWorkflowTypes": ["WatchKit", "ActionExtension"],
  "WFWorkflowInputContentItemClasses": ["WFURLContentItem"],
  "WFWorkflowActions": [
    {
      "WFWorkflowActionIdentifier": "is.workflow.actions.count",
      "WFWorkflowActionParameters": {
        "WFCountType": "Items",
        "UUID": "F292DD85-A8D2-4EBF-93E8-AC45F1C38310"
      }
    },
    {
      "WFWorkflowActionIdentifier": "is.workflow.actions.conditional",
      "WFWorkflowActionParameters": {
        "WFControlFlowMode": 0,
        "WFConditionalActionString": "0",
        "GroupingIdentifier": "51B09BBE-EF2D-4635-B820-412BADC6D64C",
        "WFCondition": "Equals"
      }
    },
    ...
    {
      "WFWorkflowActionIdentifier": "is.workflow.actions.conditional",
      "WFWorkflowActionParameters": {
        "GroupingIdentifier": "05DA8CFC-73E5-47EC-BBF6-7A23BD4D6C27",
        "WFControlFlowMode": 2
      }
    }
  ]
}

\end{verbatim}

It can be observed that this configuration file employs non-semantic hexadecimal strings to represent variables and uses keywords such as \texttt{is.workflow.actions.conditional} and \texttt{GroupingIdentifier} to implement logic controls like conditions, making it inherently difficult to read and comprehend. Consequently, we have converted it into a Python-like code format. The Python code, after transcription, variable renaming, and commenting, is shown as follows:

\begin{lstlisting}[language=Python, basicstyle=\small\ttfamily, breaklines=true]
# This line calls the function is_workflow_actions_count with a parameter of WFCountType set to 'Items', which checks the count of workflow actions related to items and assigns the result to workflow_action_count.
workflow_action_count = is_workflow_actions_count( WFCountType='Items')
# This line checks if the workflow_action_count is equal to '0', which means there are no available actions for items.
if workflow_action_count == '0':
    # If there are no actions, this line calls the function is_workflow_actions_url with a parameter of WFURLActionURL set to a specific Amazon URL to get the URL for the workflow actions and assigns it to workflow_action_url.
    workflow_action_url = is_workflow_actions_url( WFURLActionURL='https://www.amazon.com/gp/history')
    # This line displays the webpage defined by workflow_action_url by calling the is_workflow_actions_showwebpage function.
    is_workflow_actions_showwebpage( WFURL=workflow_action_url)
# This line starts the else clause that executes if 'UpdateKit' is not found in my_workflows.
else:
    # In this line, the code prompts the user for input with 'Please enter the value:', captures it, and calls the function is_workflow_actions_getvariable to get a corresponding variable and assigns the result to user_input_value.
    user_input_value = is_workflow_actions_getvariable( WFVariable=f'{input("Please enter the value:")}')
    # This line processes the user_input_value by calling the function is_workflow_actions_detect_link, which extracts a link from the user's input, and assigns the detected link to detected_link.
    detected_link = is_workflow_actions_detect_link( WFInput=user_input_value)
    # Here, the detected_link is used as input for the function is_workflow_actions_getitemfromlist to retrieve an item from a list and assigns the result to item_from_list.
    item_from_list = is_workflow_actions_getitemfromlist( WFInput=detected_link)
    # Finally, this line displays the webpage associated with the retrieved item from item_from_list by calling is_workflow_actions_showwebpage.
    is_workflow_actions_showwebpage( WFURL=item_from_list)
# This line retrieves the user's workflows by calling the function is_workflow_actions_getmyworkflows and assigns the result to my_workflows.
my_workflows = is_workflow_actions_getmyworkflows()
# This line checks if 'UpdateKit' exists in the user's workflows.
if 'UpdateKit' in my_workflows:
    # If 'UpdateKit' is found, this line creates a dictionary named updatekit_details that contains the details for the update kit, including its name, version, and RoutineHub ID.
    updatekit_details = {'Shortcut Name': 'Buy Kindle Book', 'Current Version': '1.0', 'RoutineHub ID': '1360'}
    # This line calls the function is_workflow_actions_runworkflow to execute the workflow named 'UpdateKit' with the parameters WFShowWorkflow set to False and WFInput set to the details from updatekit_details.
    is_workflow_actions_runworkflow( WFWorkflowName='UpdateKit', WFShowWorkflow=False, WFInput=updatekit_details)
    # This line contains the pass statement, indicating that if 'UpdateKit' is not found, the program will do nothing.
    pass
\end{lstlisting}

\begin{figure}[!t]
    \centering
\includegraphics[height=0.95\textheight]{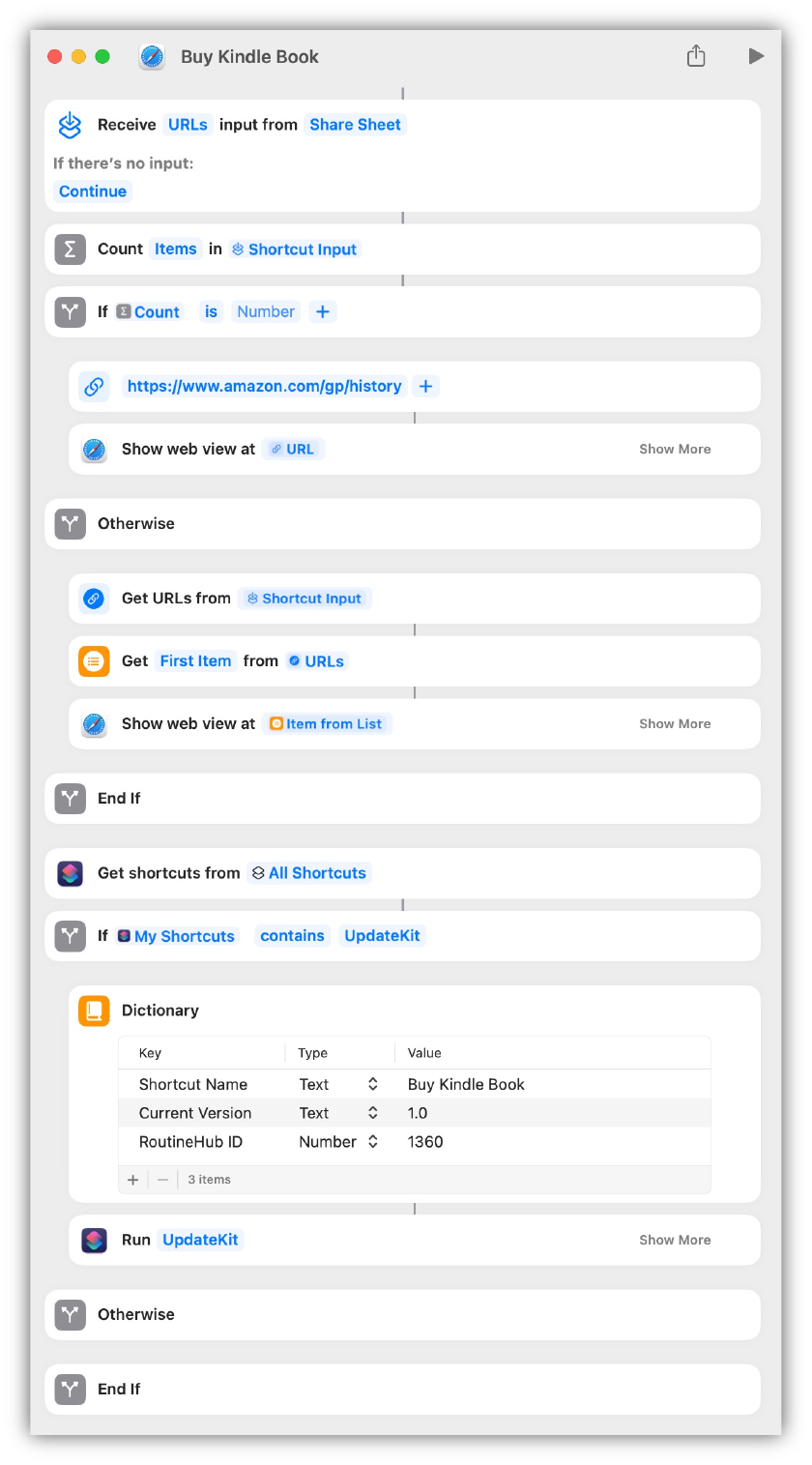}
    \caption{The visual interface of the shortcut \texttt{RoutineHub · Buy Kindle Book}.}
\label{fig:case_study_shortcut}
\end{figure}
We can clearly see that the transcribed Python code is of high quality, with strong readability, making it well-suited for training LLMs that have been pre-trained on extensive Python code.

For easy understanding, we also present the visual interface of this shortcut in Figure \ref{fig:case_study_shortcut}, which corresponds to the transcribed Python code on a line-by-line basis.

\section{Case Study of \BenchTitle} \label{case_workflow}
In this section, we provide a workflow example of \BenchTitle based on the shortcut listed in Appendix \ref{sec::case_shortcuts}. We omit the annotated workflow code that has been shown in Appendix \ref{sec::case_shortcuts}.

The query $\mathcal{Q}$ is:
\begin{lstlisting}[basicstyle=\ttfamily, breaklines=true]
Could you guide me on how to design a solution that manages workflow actions? I need it to check for existing items, retrieve purchase history from a specific source if no items are present, and also provide the option to input a specific item for detailed retrieval if items are available. Additionally, I want the solution to trigger a specific workflow if it is available within my existing processes.
\end{lstlisting}

The list of provided APIs is: 
\begin{lstlisting}[basicstyle=\ttfamily, breaklines=true]
[
    'is.workflow.actions.showwebpage', 
    'is.workflow.actions.getitemfromlist', 
    'is.workflow.actions.getvariable', 
    'is.workflow.actions.url', 
    'is.workflow.actions.getmyworkflows', 
    'is.workflow.actions.count', 
    'is.workflow.actions.runworkflow', 
    'is.workflow.actions.detect.link'
]
\end{lstlisting}

The task plan $\mathcal{P}$ is: 
\begin{lstlisting}[basicstyle=\ttfamily, breaklines=true]
1. **Start**
   - The process begins.
2. **Retrieve Workflow Action Count**
   - Call the function `is_workflow_actions_count` with parameter `WFCountType` set to 'Items'.
   - Store the result in `workflow_action_count`.
3. **Check Workflow Action Count**
   - **Decision**: Is `workflow_action_count` equal to '0'?
     - **Yes**: Proceed to step 4.
     - **No**: Proceed to step 10.
4. **Get Workflow Action URL**
   - Call the function `is_workflow_actions_url` with parameter `WFURLActionURL` set to 'https://www.amazon.com/gp/history'.
   - Store the result in `workflow_action_url`.
5. **Display Webpage**
   - Call the function `is_workflow_actions_showwebpage` with `WFURL` set to `workflow_action_url`.
6. **End Workflow Action Check**
   - End the process of checking workflow actions if the count is '0'.
7. **Retrieve My Workflows**
   - Call the function `is_workflow_actions_getmyworkflows`.
   - Store the result in `my_workflows`.
8. **Check for UpdateKit**
   - **Decision**: Is 'UpdateKit' in `my_workflows`?
     - **Yes**: Proceed to step 9.
     - **No**: Proceed to step 11.
9. **Handle UpdateKit**
   - Create a dictionary `updatekit_details` with:
     - 'Shortcut Name': 'Buy Kindle Book'
     - 'Current Version': '1.0'
     - 'RoutineHub ID': '1360'
   - Call the function `is_workflow_actions_runworkflow` with:
     - `WFWorkflowName` set to 'UpdateKit'
     - `WFShowWorkflow` set to False
     - `WFInput` set to `updatekit_details`.
10. **Receive User Input (if no update kit)**
   - Prompt user: "Please enter the value:".
   - Capture input and call the function `is_workflow_actions_getvariable`.
   - Store the result in `user_input_value`.
11. **Detect Link from Input**
   - Call the function `is_workflow_actions_detect_link` with `WFInput` set to `user_input_value`.
   - Store the result in `detected_link`.
12. **Get Item from List**
   - Call the function `is_workflow_actions_getitemfromlist` with `WFInput` set to `detected_link`.
   - Store the result in `item_from_list`.
13. **Show Item Webpage**
   - Call the function `is_workflow_actions_showwebpage` with `WFURL` set to `item_from_list`.
14. **End Process**
   - The process concludes after performing the respective actions based on the conditional checks.
\end{lstlisting}

\section{Limitations}

While the framework proposed in this paper represents notable progress in workflow orchestration, it also has certain limitations that warrant discussion. First, the APIs used in our work are exclusively derived from Apple Shortcuts application, resulting in a lack of coverage across more diverse fields and potentially limiting the generalizability of the dataset to broader application contexts. Second, our approach lacks evaluation through actual execution. This limitation arises due to the complexities involved in executing workflows, such as the need for intricate user registration and permission acquisition. Moreover, the APIs are subject to frequent changes as applications continue to evolve, making it challenging to implement a consistent executable evaluation. Consequently, our evaluation is limited to static analysis.

\section{Ethical Statement}

In this study, the dataset construction process was fully automated using LLMs and algorithms for data annotation, eliminating the need for human annotators and thereby avoiding concerns related to annotator compensation and working conditions. The data utilized was collected through web scraping from publicly accessible sources, with strict adherence to the Terms of Service~(ToS) of the respective websites. Scraping was avoided on platforms where such activity is explicitly prohibited, ensuring compliance with ethical standards. Additionally, no personally identifiable information~(PII) or private user data was collected at any stage of the research process. All data was anonymized to protect privacy and mitigate any potential ethical concerns related to user information.

\end{document}